# The phase of Hidden Momentum in Aharonov-Bohm solenoid Effect

Y.Ben-Aryeh

Physics Department, Technion-Israel Institute of Technology, Haifa, 32000 Israel

e-mail: phr65yb@ph.technion.ac.il

It is shown that the phase of the 'hidden momentum' in Aharonov-Bohm (AB) solenoid effect is equal in magnitude to the phase of the electron but with opposite sign. The phase of the hidden momentum is equal to that obtained by the energy of interference calculated in our previous paper (J.Opt.Soc.Am. B **17**, 2052, 2000).

## 1. Introduction

Aharonov-Bohm (AB) have shown [1] that electron moving in a closed circuit around a long solenoid would obtain a quantum phase shift proportional to the magnetic flux enclosed in the solenoid, although the electron moves in a region that is free of electric and magnetic fields. According to AB explanation, while classically the equations of motion are described by classical fields in Quantum-Mechanics (QM) the potentials are the essential parameters, so that the quantum phase is obtained by taking into account the vector potential. This AB effect has been verified later experimentally [2,3]. This interpretation is still under certain controversies (see e.g. [4]).

In a previous paper [5] we have shown that during the motion of the electron around the solenoid there are momentum exchanges between the mechanical solenoid and the electromagnetic (EM) field. The integral of momentum of interference over the trajectory of the electron, which is equal to the integral of the energy of interference over the corresponding time, gives the AB effect.

Due to the above alternative calculation of AB phase the author has suggested in a previous publication [6] to investigate three-body forces in AB systems, but following previous claims the solenoid does not accelerate due to the effect of 'hidden momentum' [7]. I would like to show here that our phase calculation [5] corresponds to the calculation of the 'hidden momentum' phase which is equal in magnitude to that of AB phase but with opposite sign. The interpretation for such phase relation is given in the present Letter. Although the analysis is shown for the AB solenoid effect it might have implications also to other AB effects.

**2. The relations between the 'hidden momentum' , the energy of interference and AB phase**

While there is no force on the electron in AB solenoid effect the electron exerts <u>two forces</u> on external systems : 1) On the mechanical solenoid it exerts the force $\vec{F}_j$ where

$$\vec{F}_j = -\frac{d}{dt}\left(\frac{q}{c}\vec{A}\right) \quad . \tag{1}$$

Here $\vec{F}_j$ refers to the force that the electron exerts on the solenoid currents, to first order in v/c, q is the electron charge and $\vec{A}$ is the vector potential. 2) Due to the magnetic moments produced by the currents in the solenoid there is an additional force related to the 'Hidden Momentum' [7] which is equal in magnitude to that of Eq. (1) but with opposite sign.

The crucial point in our previous analysis [5] was that the force on the solenoid is equal in magnitude to the rate of change of momentum of the EM field, such that we get the equality

$$\left(\frac{q}{c}\right)\vec{A}(\vec{r}) = \left(\frac{1}{4\pi c}\right)\int \left[\vec{E}_{el}(\vec{r},\vec{r}\,') \times B_{Sol}(\vec{r}\,')\right]d^3r' \qquad (2)$$

Here, $\vec{E}_{el}(\vec{r},\vec{r}\,')$ represents the electric field produced at point $\vec{r}\,'$ by electron located at point $\vec{r}$. $B_{Sol}(\vec{r}\,')$ is the magnetic field at point $\vec{r}\,'$, assumed to be constant in the solenoid. So we find that the integral of Eq. (2) represents the momentum of interference for the electric field of the electron in the solenoid and the solenoid magnetic field. By using the identity of Eq. (2) we find the following relation for the AB electron phase

$$\Delta\phi = \left(\frac{q}{c\hbar}\right)\oint \vec{A}(\vec{r})\cdot d\vec{r} = \left(\frac{1}{4\pi c\hbar}\right)\oint \left[\int \vec{E}_{el}(\vec{r},\vec{r}\,') \times B_{Sol}(\vec{r}\,')d^3r'\right]\cdot d\vec{r} \qquad (3)$$

The integral in the square brackets of Eq. (3) represents the momentum of interference for the electric field of the electron in the solenoid and the solenoid magnetic field. By simple transformations of Eq.(3) we obtained an alternative equation for AB solenoid phase given by [5]

$$\Delta\phi = \frac{1}{\hbar}\int\left\{\int \left[\frac{\vec{B}_{el}(\vec{r},\vec{r}\,')\cdot B_{Sol}(\vec{r}\,')}{4\pi}\right]d^3r'\right\}dt \qquad (4)$$

The expression in braces of Eq. (4) represents the energy of interference where in the derivation of Eq. (4) we have transformed the triple product of Eq. (3) and used the relations:

$$\vec{B}_{el}(\vec{r},\vec{r}\,') = \left[\vec{v}_{el}(\vec{r})/c\right] \times \vec{E}_{el}(\vec{r},\vec{r}\,') \quad ; \quad d\vec{r} = \vec{v}dt \quad . \qquad (5)$$

The integration in Eq. (4) is over the volume of the solenoid and the time t that it takes the electron to go around the solenoid. We have calculated the complicated integrals of Eq. (4) and have shown [5] that this equation gives the same magnitude of phase as that obtained by AB phase, but with opposite sign.

It has been shown in previous studies that a magnetic moment $\vec{\mu}$ in an electric field $\vec{E}_0$ acquires a momentum [7] :

$$\vec{P} = \frac{1}{c} \vec{\mu} \times \vec{E}_0 \quad , \qquad (6)$$

where this momentum is often referred as 'hidden momentum' [7]. In the present Letter I would like to show that the expression on the right side of Eq. (2) represents the 'hidden Momentum' in the solenoid when we integrate over all the magnetic moments produced by the solenoid currents. In order to show this assertion we can use the following calculation.

Let us take into account that the solenoid current is composed of many circular loops of the same radius all carrying the same current. A plane circuit which encloses the plane area $\vec{S}$, acts at great distances like a permanent magnet of magnetic moment $\vec{\mu} = i\vec{S}/c$ where the direction of $\vec{S}$ is perpendicular to this plane. For our purpose we divide the surface $\vec{S}$ into infinitesimal area elements $d\vec{S}$ each having infinitesimal magnetic moment

$$d\vec{\mu} = id\vec{S}/c \qquad . \qquad (7)$$

The proof of Eq. (7) follows by using a quite common procedure. We divide the plane surface $\vec{S}$ bounded by the current loop into elements of area $d\vec{S}$ where all elements $d\vec{S}$ have the same direction along the solenoid axis. For all such surface elements we assume the same current $i$ along the circuit enclosing the surface elements $d\vec{S}$. The contribution arising from the common boundary of two elements (e.g. $d\vec{S}_1$ and $d\vec{S}_2$) exactly cancel each other , since they always occur in pairs, and that with opposite sign. Thus after addition we are left only with integral over the original boundary contour of the large surface $\vec{S}$. The use of Eq. (7) is essential for exact distance relation between the electron location and the magnetic moment infinitesimal element.

Assuming that the number of current loops for a unit distance along the solenoid is given by *n* then the magnetic moment element for a distance *dl* along the solenoid axis is given by

$$d\vec{M} = \frac{ni}{c} dl\, d\vec{S} \qquad . \qquad (8)$$

The magnetic field intensity inside a very long solenoid is given by

$$\vec{B} = ni\, d\hat{S} \quad , \qquad (9)$$

where $d\hat{S}$ is a unit vector along the solenoid axis. Then we get

$$d\vec{M} = \frac{\vec{B}}{c} dV \qquad . \qquad (10)$$

Substituting $d\vec{M} = \vec{\mu}$ in Eq. (7) we find with a more explicit notation

$$d\vec{P} = \frac{1}{c}\left[\vec{E}_{el}(\vec{r},\vec{r}\,') \times B_{Sol}(\vec{r}\,') d^3r'\right] \quad , \qquad (11)$$

which is equal to the integrand of Eq. (2). As Eq. (2) is our basic equation for deriving the alternative expression for AB phase our further derivations represented by Eqs. [3-5] can be followed from the use of Eq. (11) and referred to as a quantum calculation of the 'Hidden Momentum Phase'.

The result that we have two phases that of the vector potential $\vec{A}$ and that of the 'Hidden Mometum' follows from the momentum conservation principle. The vanishing of the sum of these two phases might be related to a certain complementary principle [8]. The idea that on the solenoid two classical forces are operating which are equal in magnitude but with opposite sign has been implemented also in Ref. [9].

## 3. Discussion and Conclusion

According to Newton Third Law whenever two bodies interact the force $\vec{F}_{2,1}$ that body 1 exerts on body 2 is equal in magnitude and opposite in sign to the force $\vec{F}_{1,2}$ that body 2 exerts on body 1 (Action and Reaction principle). When we consider the solenoid as one entity the sum of the forces due to the vector potential and that due to the 'Hidden Momentum' vanish so that the solenoid does not accelerate. In order to observe separately the effects of these two forces we need a measurement process that will distinguish between these two forces. Such method of measurement for the ordinary solenoid system becomes very difficult. However, one should take into account that the above two classical forces follow from the action of the third body- the electron, although the electron does not suffer any classical force. Under a shielding of the solenoid [3] the electric field of the electron does not penetrate into the solenoid and Eq. (2) is not valid as the right side of Eq. (2) vanish while the vector potential given on the left side of this equation would still lead to the AB phase [3]. Under such conditions I expect that the vector potential $\vec{A}(\vec{r})$ would lead to classical forces between the electron and the induced currents on the metallic shielding envelope. I expect also that in this case there will be an additional force which will satisfy the conservation of momentum requirement. The two phases following from the two forces in the ordinary solenoid system can, however, be separated by interference experiments for the electron as in such experiments the electron phase is affected only by the vector potential.

Although a very special system has been treated in the present letter there is a fundamental aspect following from the present analysis. A three-body-system in which a system A (e.g. the electron) affects both system B (e.g. the mechanical solenoid) and system C (e.g. the hidden EM field momentum) while system A does not suffer any classical force might be important as a new basic three-body-effect in QM.

**Acknowledgement**

The author would like to thank Prof. Lev Vaidman for helpful discussions